\documentstyle[12pt,epsfig]{article}

\begin{document}

\title{Exploring the Gamma Ray Horizon with the next generation of
Gamma Ray Telescopes. Part 2: Extracting cosmological parameters
from the observation of $\gamma$-ray sources}
\author{O.Blanch, M.Martinez\\
{\it IFAE, Barcelona (Spain) }}

\maketitle

\begin{abstract}
The potential of the new generation Cherenkov Telescopes to
measure the energy spectrum of both, the already established
extragalactic very high energy $\gamma$-ray emitters and the best
very high energy candidates from the EGRET catalogue is discussed.
By a realistic simulation of the analysis of the expected
extrapolated energy spectra, it is shown that the foreseen
capability and precision of these instrument to measure the Gamma
Ray Horizon may open the door to competitive measurements of the
cosmological parameters.
\end{abstract}


\newpage


\section{Introduction}

Imaging Air \v{C}erenkov Telescopes (IACT) have proven to be the
most successful tool developed so far to explore the $\gamma$-ray
sky at energies above few hundred GeV. A pioneering generation of
installations has been able to detect a handful of sources and to
start a whole program of very exciting physics studies. Nowadays a
second generation of more sophisticated Telescopes is starting to
provide new  observations. One of the main characteristics of some
of the new Telescopes is the potential ability to reduce the gamma
ray energy threshold below $\sim 30$ GeV \cite{MAGIC}.

In the framework of the Standard Model of particle interactions,
high energy gamma rays traversing cosmological distances are
expected to be absorbed through their interaction with the diffuse
background radiation fields, or Extragalactic Background Field
(EBL), producing $e^+ e^-$ pairs. Then the flux is attenuated as a
function of the gamma energy $E$ and the redshift $z_{q}$ of the
gamma ray source. This flux reduction can be parameterized by the
optical depth \(\tau(E,z_{q})\), which is defined as the number of
e-fold reductions of the observed flux as compared with the
initial flux at $z_{q}$. This means that the optical depth
introduces an attenuation factor \(\exp[-\tau(E,z_{q})]\)
modifying the gamma ray source energy spectrum.

The optical depth can be written with its explicit redshift and energy
dependence \cite{Stecker-96} as:

\begin{equation}
\tau(E,z) =
\int_{0}^{z}dz'\frac{dl}{dz'}\int_{0}^{2}dx \,
\frac{x}{2}\int_{\frac{2m^{2}c^{4}}{Ex(1+z')^{2}}}^{\infty}
d\epsilon\cdot n(\epsilon,z') \cdot \sigma[2xE\epsilon(1+z')^{2}]
\label{eq:OpD}
\end{equation}

where $x \equiv 1-\cos\theta $ being $\theta$ the angle between the
photon directions, $\epsilon$ is the energy of the EBL
photon and $n( \epsilon ,z')$ is the spectral density at the given z'.

For any given gamma ray energy, the Gamma Ray Horizon (GRH) is
defined as the source redshift $z$ for which the optical depth is
$\tau(E,z) = 1$.

In a previous work \cite{part1}, we discussed different
theoretical aspects of the calculation of the Gamma Ray Horizon,
such as the effects of different EBL models and the sensitivity of
the GRH to the assumed cosmological parameters.

In this work we estimate with a realistic simulation the accuracy
in the determination of the GRH that can be expected from the
observation of the known extragalactic sources which will be, for
sure, studied by the new IACTs. These sources are:

\begin{itemize}
\item{} on the one hand the very high energy gamma-ray emitting
blazars already studied with the previous generation of
instruments, whose spectra will be measured now up to lower
energies, providing then a better lever arm for the determination
of the absorption cut-off energy and, \item{} on the other hand
the EGRET blazars which, by carefully extrapolating the measured
spectrum, can be expected to have a very high energy tail, and
that very likely haven't been observed at the previous generation
of instruments due to the effect of the absorption cut-off.
\end{itemize}

The above sources, are distributed in a broad range of redshifts
and therefore the measurement of their GRH provides a nice mapping
of the GRH as a function of redshift, which we use to constraint
our understanding on the EBL density and on the cosmological
parameters entering the prediction of the $GRH(z)$.

 It is worth to point out that besides these "bread and
butter" sources, for which our assumptions are quite
plausible, the reduction of the energy threshold is expected
to allow the new instruments to discover a plethora of new sources
which has been hidden from our observation so far. An important
part of this new population is expected to be at relative large
redshift ($z>2$) where a big fraction of galaxies were AGNs and
hence could produce very high energy gamma-rays. These new sources
could drastically improve the results discussed here which,
therefore, have to be considered as being rather conservative.

The outline of this work is the following: first we discuss the
choice of the sources used and the way the spectra have been
extrapolated. In section three we describe the procedure used to
fit the expected spectra to the GRH and we discuss possible
systematic uncertainties. Section four deals with the fit of the
redshift dependence of the GRH to the cosmological parameters and
the discussion on the systematic uncertainties due to our poor
knowledge of the EBL density. Finally in section 5 we summarize
the results obtained and explain our conclusions.

\section{Flux extrapolation}
\label{extrapolation}

Although most of the conclusions of this work should apply up to
to a large extent to any low-threshold installation, the specific
study we present here assumes an installation on the Northern
Hemisphere with a zenith threshold of about $30 GeV$ such as the
MAGIC Telescope.

For the first observations of any new installation, priority will
be given to the investigation of the well-established
extragalactic $\gamma$--ray sources which will allow to
cross-check measurements with other experiments\cite{flix-ICRC03}.
The well-established extragalactic sources observed in the
Northern Hemisphere are listed in Tab.\ref{tab:known_sources}.

\begin{table}
\begin{center}
\fontsize{9}{10} \selectfont
\begin{tabular}{|lccc|}
\hline
 Source  & Redshift & $f_{o}$ & $\alpha$ \\
         & $z$ & [$10^{-11} \gamma cm^{-2} s^{-1} TeV^{-1}$] & [] \\
\hline
 Mkn~421      & 0.031    & 12.1 & 2.18          \\
 1ES~1426+428 & 0.129    & 0.2  & 2.6           \\
 Mkn~501      & 0.034    & 10.8 & 1.92           \\
 1ES~1959+650 & 0.047    &    -    &      -         \\
 PKS~2155-304 & 0.116    &    -    &      -         \\
 1ES~2344+514 & 0.044    &    -    &      -         \\
\hline
\end{tabular}
\end{center}
\caption{\label{tab:known_sources}BL Lac objects observed
in the $TeV$ band. The absolute flux and the spectral index has been
taken from flaring periods (\cite{Aharonian-02a,Aharonian-02b,Aharonian-00})}
\end{table}

Apart from these few $\gamma$-ray sources detected by \v Cerenkov
Telescopes, the observation program will unavoidably include
looking for new extragalactic $\gamma$-ray emitters. For that, one
of the most plausible approaches that can be followed is based on
the third EGRET catalogue \cite{Hartman-99}. It gives us the most
complete and recent experimental situation for extragalactic
sources at the highest satellite energies (from 100 $MeV$ up to 10
$GeV$). A suitable set of blazar candidates for a MAGIC-like
installation can be obtained by extrapolating the Northern
Hemisphere EGRET blazars fitted spectra to the MAGIC energy
detection range\cite{flix-ICRC03}.\par

Unfortunately the extrapolation to the MAGIC energies, both from
well-stablished $\gamma$-ray sources and EGRET candidates, is not
straight-forward since several phenomena have to be taken into
account:

\begin{itemize}

\item{} On the one hand the $\gamma$--ray absorption due to pair
production by the EBL, where the uncertainty is sizeable due to
the poor knowledge of the EBL.

\item{}On the other hand, in the frame of leptonic synchrotron
self-Compton (SSC) emission models, the Inverse Compton spectral
break will be, for some sources, between the EGRET and MAGIC
energies. With the current data, for some of the sources it is
difficult to predict the energy of the spectral break with better
precision than 2 orders of magnitude which, because of the sharp
$\gamma$-ray spectra, may reflect into several orders of magnitude
in the extrapolated flux above that energy. Moreover, these
$\gamma$-ray candidates are mainly supposed to be AGN, which may
have large time-variations in their flux.\par

\end{itemize}

\begin{figure}[h]
  \begin{center}
    \includegraphics[width= 32.0pc, height=12.0pc]{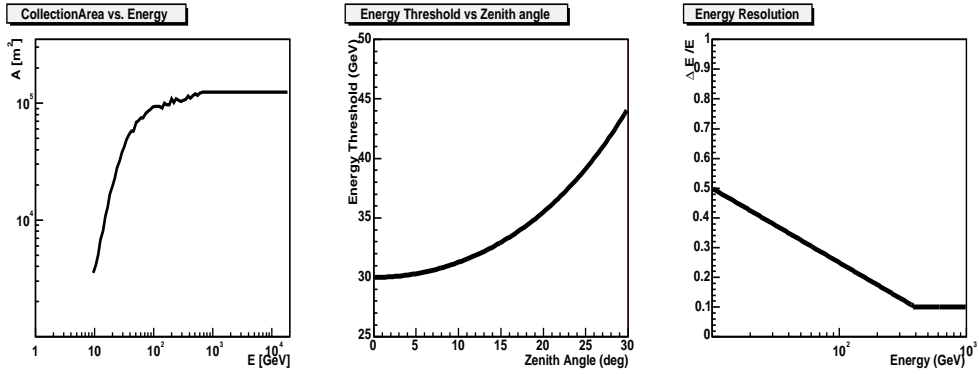}
  \end{center}
  \vspace{-0.5pc}
  \caption{Assumed Telescope characteristics. Efficient Trigger
    Collection Area for a CT similar to the
    MAGIC dimensions and performance. Energy threshold as a function of
    zenith angle as measured by previous CT, and energy resolution
    form MAGIC TDR \cite{MAGIC}}
\label{fig:magic}
\end{figure}

For these reasons, several hypotheses are needed for the
extrapolation of the flux:

\begin{itemize}

\item{} We assume that all sources are observed in a high flaring
state. At first glance this may seem a very optimistic hypothesis
but it is important to point out that MAGIC, and the rest of the
second generation of Cerenkov Telescopes, are expected to have a
large amount of source candidates to be observed and a flaring AGN
will normally be a target of opportunity.

\item{} Some assumptions about the Inverse Compton spectral break
and the spectral index are also needed. Following the strategy
that some AGN search study groups propose for the next generation
of telescopes \cite{flix-ICRC03}, if the spectral index measured
by EGRET is larger than $2.0$, we just extrapolate to higher
energies with that spectra. If it is smaller than $2.0$ we assume
the spectral break at around 50 GeV and a $2.4$ index after
that,using\cite{Smith}:
\begin{equation}
\frac{dN}{dE}=\frac{f_{0}E^{-\alpha}}{(1+(\frac{E}{E_{b}})^{f})^{\beta/f}}
\end{equation}
where $\alpha + \beta = \gamma$, $f$ can go from $1$ to $2.3$ (we
use $2.0$) and $E_{peak}=E_{b}((2-\alpha)/(\gamma-2))^{1/\gamma}$

\item{} We assume an specific model for the relevant EBL
\cite{Kneiske}, which will lead to a set of given $GRH(z)$ values.

\item{} Several characteristics of the actual IACT need to be
assumed. To be more precise, the effective collection area as a
function of the gamma-ray energy, the energy threshold as a
function of zenith angle and the energy resolution, inspired on
the MAGIC Telescope simulations are assumed (see figure
\ref{fig:magic}).

\end{itemize}

With all these hypotheses, we have simulated the flux spectrum for
each source assuming $50$ hours observation time and we have
determined the actual precision with which it might be measured.
For that the following steps have been done:

\begin{itemize}

\item{} We get the $dN/dE$ emitted by the source using the
extrapolation described above.

\item{} The Optical Depth is applied using the EBL model detailed
in \cite{Kneiske} to get the $dN/dE$ reaching the Earth.

\item{} The effective collection area as a function of the
gamma-ray energy is used to get the $dN/dE$ detected by the
telescope.

\item{} Binning in energy, the number of gamma-rays is computed.
This number is used to calculate the statistical error using as an
error the square root of gammas. A multiplicative factor to the
sqrt($\gamma$) is applied to estimate the final error including the
background and the telescope behaviour. This factor has been extracted
from data coming from the first generation of Cerenkov Telescopes
\cite{Aharonian-02a}, hence we assume an understanding of the
background and detector at the level achieved by them.

\item{} Using the energy resolution and the number of gamma-rays
per bin, we compute the energy uncertainty. Later on, this energy
uncertainty is taken into account in the fit of the spectrum.

\end{itemize}

\begin{figure}[h]
  \begin{center}
    \includegraphics[height=18pc]{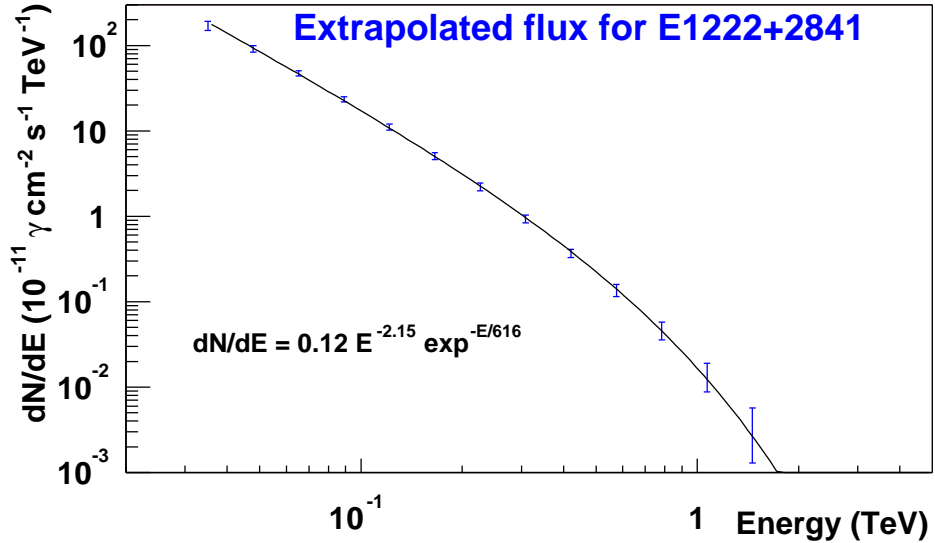}
  \end{center}
  \vspace{-0.5pc}
  \caption{Extrapolated flux for source E1222+2841}
\label{fig:flux}
\end{figure}

Figure \ref{fig:flux} shows the extrapolated flux for the EGRET
source EG1222+2841 assuming $50$ hours observation time and the
best fit to the spectral index and the spectrum cut-off energy
(assuming a simple analytical expression that will be justified
below) after the above steps have been completed.

\section{Gamma Ray Horizon fits}

The differential flux seen at the earth should actually be:

\begin{equation}
\frac{dN}{dE}= ( \frac{dN}{dE} )_{unabsorbed}\cdot e^{-\tau(E,z)}
\end{equation}

The actual dependence of $\tau$ on $E$ and $z$ is quite complex
and cannot be cast on a simple analytical expression \cite{part1}.
Nevertheless, in first approximation, the exponential suppression
term can be expressed as and e-fold reduction in the energy
$E_{0}$, which will coincide with the Gamma Ray Horizon energy.
Moreover, in case the threshold energy of the Telescope is well
below the spectrum cut-off energy, the emitted flux $dN/dE$ can be
well approximated for most of the sources by a power law.
Therefore, in first approximation one gets an analytical
expression that can be used to fit the spectrum and get the energy
of the GRH ($E_{0}$):

\begin{equation}
\frac{dN}{dE}=f_{0}\cdot E^{-\alpha} \cdot e^{-(\frac{E}{E_{0}})^{\beta}}
\end{equation}

It is true that at the energy range that MAGIC will reach one can
argue that in some cases the effect of the Inverse Compton
spectral break may play a role. Actually as it seen in the section
\ref{extrapolation} we took into account that, whenever it was
needed. Unfortunately the expected energy threshold for MAGIC is
so close to the assumed break energy that there is not lever arm
to get information on that with the assumed $50$ hours observation
time. Therefore, the previous equation has been used to fit the
spectrum and get the GRH energy for all the extrapolated sources.
The error due to this simplification has been included in the
systematics.

The result and precision of the GRH energy fitted for every source
as well as the theoretical predictions are shown in the table
\ref{tab:GRH}.

\begin{table}[t]
\begin{center}
\fontsize{9}{10} \selectfont
\begin{tabular}{|c|ccc|}
\hline
Source Name & z & $E_{0} (GeV)$ & $\sigma_{E_{0}} (GeV)$ \\
\hline \hline
Mrk 421 , 3EG J1104+3809 & 0.031 & 5203  & 448 \\
W Comae , 3EG J1222+2841 & 0.102 & 615.7 & 23.9 \\
         3EG J1009+4855 & 0.200   & 355.2 & 2.6 \\
OJ+287 , 3EG J0853+1941 & 0.306 & 255.3 & 9.1 \\
4C+15.54 , 3EG J1605+1553 & 0.357 & 224.3 & 7.7 \\
        3EG J0958+6533 & 0.368 & 219.1 & 15.2 \\
        3EG J0204+1458 & 0.405 & 201.7 & 12.7 \\
        3EG J1224+2118 & 0.435 & 189.1 & 21.6 \\
3C 279  , 3EG J1255-0549 & 0.538 & 155.4 & 1.3 \\
        3EG J0852-1216 & 0.566 & 148.0 & 2.6 \\
4C+29.45 , 3EG J1200+2847 & 0.729 & 114.4 & 3.2 \\
CTA026  , 3EG J0340-0201 & 0.852 & 96.61 & 1.76 \\
3C454.3 , 3EG J2254+1601 & 0.859 & 95.74 & 0.73 \\
        3EG J0952+5501 & 0.901 & 90.82 & 7.8 \\
        3EG J1733-1313 & 0.902 & 90.71 & 4.8 \\
OD+160  , 3EG J0237+1635 & 0.940  & 86.68 & 1.00 \\
        3EG J2359+2041 & 1.070 & 75.31 & 7.4 \\
        3EG J0450+1105 & 1.207 & 66.58 & 3.8 \\
        3EG J1323+2200 & 1.400 & 57.87 & 2.4 \\
        3EG J1635+3813 & 1.814 & 46.69 & 1.8 \\
\hline
Mrk 501 & 0.034 & 4274 & 115 \\
1ES J1426+428  & 0.129 & 504.4 & 61.1 \\
\hline

\end{tabular}
\caption{\label{tab:GRH} GRH fit predictions for the 22 sources
considered.}
\end{center}
\end{table}

\section{Cosmology}

As discussed already in \cite{part1}, from the expression of the
Optical Depth (equation \ref{eq:OpD}), it is clear that some
fundamental cosmological parameters such as the Hubble constant
and the cosmological densities play an important role in the
calculation of the GRH, since:

\begin{equation}
\frac{dl}{dz}=c\cdot\frac{1/(1+z)}{H_{0}[\Omega_{M} (1+z)^{3}
+ \Omega_{K} (1+z)^{2} + \Omega_{\Lambda}]^{1/2}}
\label{eq:dldz}
\end{equation}

Therefore the measurement of the GRH for sources at several
redshifts will open the possibility to obtain constraints in some
fundamental cosmological parameters \cite{part1}.

Conceptually the measurement of the GRH as a function of the
redshift, provides a new distance estimator which has the
following features:

\begin{itemize}

\item{}It is independent and behaves differently from the
luminosity-distance relation currently used by the Supernovae 1A
observations (see \cite{part1}).

\item{}It does not rely on the existence of a time-independent
standard-candle as do the Supernovae 1A measurements, although it
relies on the existence of a cosmological infrared EBL which, in
first approximation, is assumed to be uniform and isotropic at
cosmological scales.

\item{}It uses Active Galactic Nuclei as sources, and therefore
may allow the study of the expansion of our universe up to the
highest observable redshifts. In this sense this method might
complement the picture provided by Supernovae 1A exploring the
farthest universe.

\end{itemize}

\begin{figure}[h]
  \begin{center}
    \includegraphics[height=18pc]{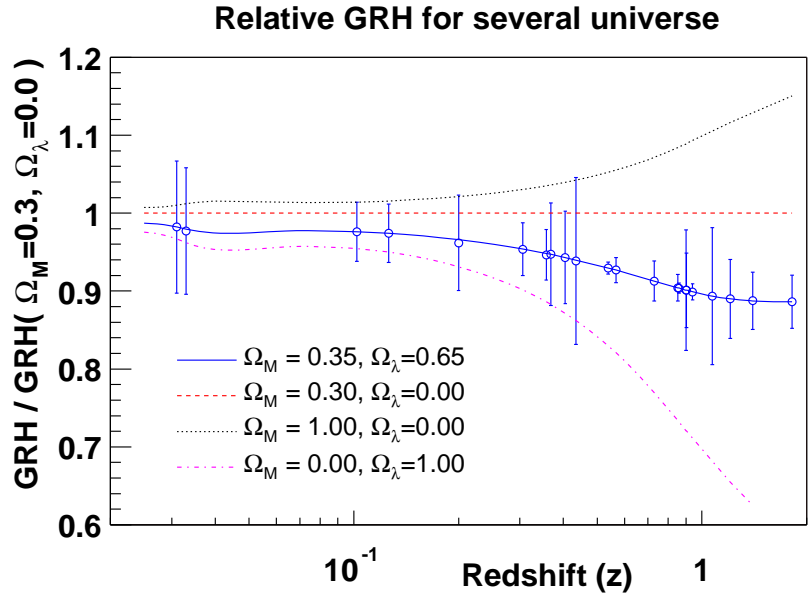}
  \end{center}
  \vspace{-0.5pc}
  \caption{Simulated GRH measurements and the GRH predictions
  for different universe models}
\label{fig:models}
\end{figure}

In figure \ref{fig:models}, the simulated GRH energy measurements
and their estimated expected uncertainties are plotted together
with the theoretical GRH predictions for several extreme
hypothetical universes.\par

The prediction of the GRH as a function of the redshift has
basically only the following parameters: the Hubble constant, the
cosmological densities and last, but not least, the EBL density
spectrum as a function of redshift. Assuming the later perfectly
known, one can try to use the simulated measurements of table
\ref{tab:GRH} to fit the cosmological parameters. This is a
four-parameter fit which, if tried with "brute-force" turns out to
be inviable. Instead we've followed the strategy developed in
\cite{martinez-miquel} for a similar problem, which consists on
the use of a multi-dimensional interpolating routine based upon
the algorithms of \cite{NumericalRecipesFortran}. We have checked
that within the parameter intervals relevant for the fits
discussed in this work, that interpolation produces results which
reproduce the exact predictions with the required accuracy.

\begin{figure}[h]
  \begin{center}
    \includegraphics[height=18pc]{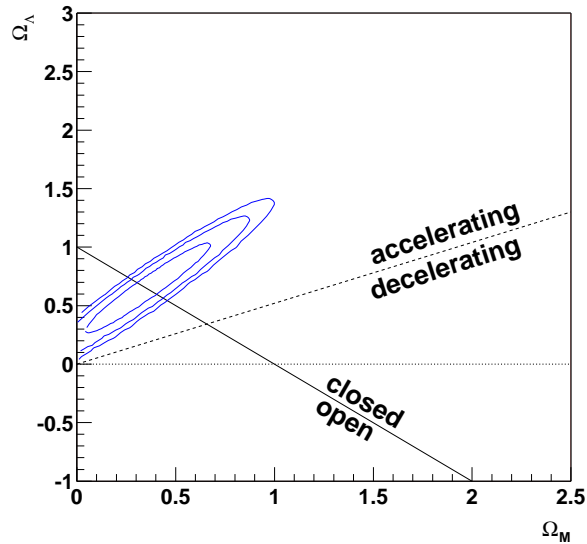}
  \end{center}
  \vspace{-0.5pc}
  \caption{Expected contour levels of $68 \%$, $95\%$ an $99\%$ for
    the $\Omega_{M}-\Omega_{\lambda}$ plane based on the GRH.}
\label{fig:contours}
\end{figure}

In Figure \ref{fig:contours}, the $\Delta \chi^2=2.3$, $5.99$ and
$9.21$ contours, corresponding to a two-parameter confidence areas
of $68 \%$, $95\%$ an $99\%$ respectively in the
$\Omega_{M}-\Omega_{\lambda}$ plane are plotted. Despite the GRH
measurements allow also to obtain some information on the Hubble
constant, experiments looking to a closer distance can do it much
better. For this reason the best fit confidence region has
been computed assuming an external constrain in $H_{0}$ of 4.0 Km
s /Mpc \cite{Spergel-03}.\par

These contours can be compared with the results from the
combination of all Supernovae 1A measurements at relatively low
redshift \cite{supernovae} to show that there is roughly a factor
2 improvement in the expected uncertainties. Actually, these
contours have a size similar to the one recently claimed by the
use of very distant Supernovae \cite{Raux}, which shows the
 need for looking at large redshifts (1-2) to improve in the measurement
 of  $\Omega_{M}$ and
 $\Omega_{\lambda}$. It is also worth to point out that it will
also be a significant measurement taking into account other
current techniques \cite{WMap,GalaxyCounting}, since the explored
parameter space is rather orthogonal.

\subsection{Systematics}

So far mostly statistical uncertainties have been taken into
account. This could be unrealistic since large systematical
uncertainties could eventually appear in some of the assumptions
taken.

Two kind of systematic uncertainties have been taken into account:

\begin{itemize}

\item{} On the one hand, "experimental" systematics, which we
believe will be dominated by the global energy scale, which enters
directly in the GRH determination and is not very well known in
Cherenkov Telescopes. We have assumed a conservative $15 \%$
global energy scale systematic uncertainty.

\item{} On the other hand, "theoretical" systematics. As already
stated, the GRH behavior with the redshift does not depend only on
the cosmological parameters but also on the EBL density as a
function of redshift assumed. The fact that, at present, the EBL
is not well measured at the relevant energy range, and its
redshift dependence is not well known, forces the use of different
models which differ substantially in their predictions
\cite{COBE-IV,Kneiske}. The assumption of different EBL models is
expected to produce important uncertainties in the determination
of these cosmological parameters.

\end{itemize}

\begin{figure}[h]
  \begin{center}
    \includegraphics[height=18pc]{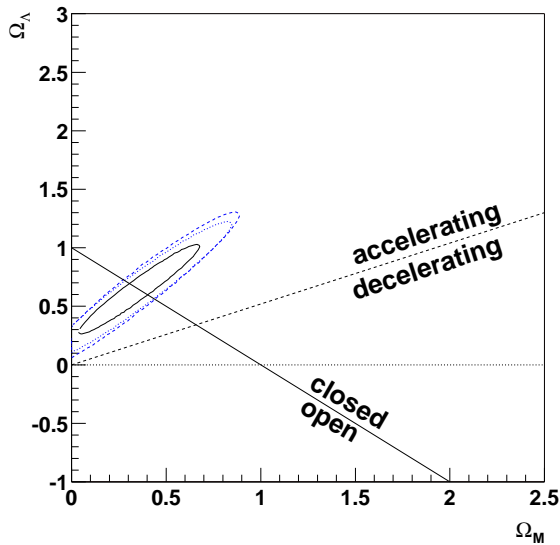}
  \end{center}
  \vspace{-0.5pc}
  \caption{Contour levels of $68 \%$ for the
    $\Omega_{M}-\Omega_{\lambda}$ plane showing the effect of including
    different "experimental" systematics. Solid line is the contour with
    only statistical errors. Dotted line includes the
    error due to the fit simplification. Finally, in the dashed
    line it is also included the systematic due to the energy scale.}
\label{contours-experimental-systematics}
\end{figure}

For what concerns the "experimental" systematics, figure
\ref{contours-experimental-systematics} illustrates the modest
size of the estimated effect due to the effect of the global
energy scale and the simplifications done in the spectral fit.

\begin{figure}[h]
  \begin{center}
    \includegraphics[height=18pc]{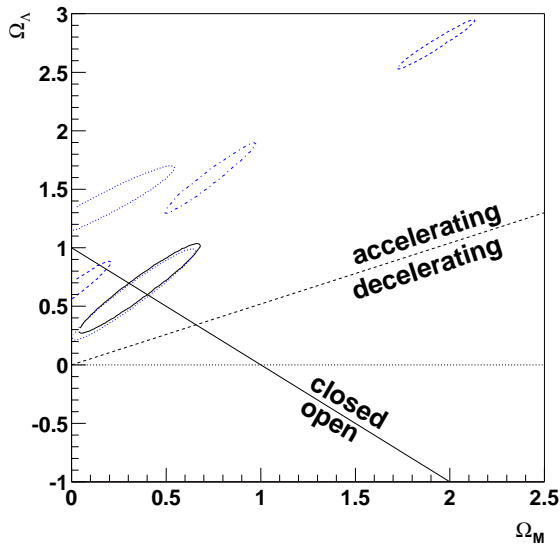}
  \end{center}
  \vspace{-0.5pc}
  \caption{Contours levels of $68 \%$ for
    several EBL models.}
\label{contours-EBL-models}
\end{figure}

For what concerns the "theoretical" systematics, the situation is
more complex. To estimate them, we have used a set of models
\cite{Kneiske03} which are somehow representative of the different
approaches followed up to now for the prediction of the EBL and that,
so far, have not been excluded by the existing relevant
observations. Each model has a rather complex set of assumptions and
physics ansatzs based upon some observations and therefore it does not
look feasible to parameterize all them in a simple manner which would
allow us to fit them to the GRH data. The contours obtained for the
different EBL models are shown in figure
\ref{contours-EBL-models}. Nevertheless, there are a couple of
facts that can be taken into account to try to obtain a plausible
estimate of their effect in the cosmological parameter fit.

\begin{figure}[h]
  \begin{center}
    \includegraphics[height=18pc]{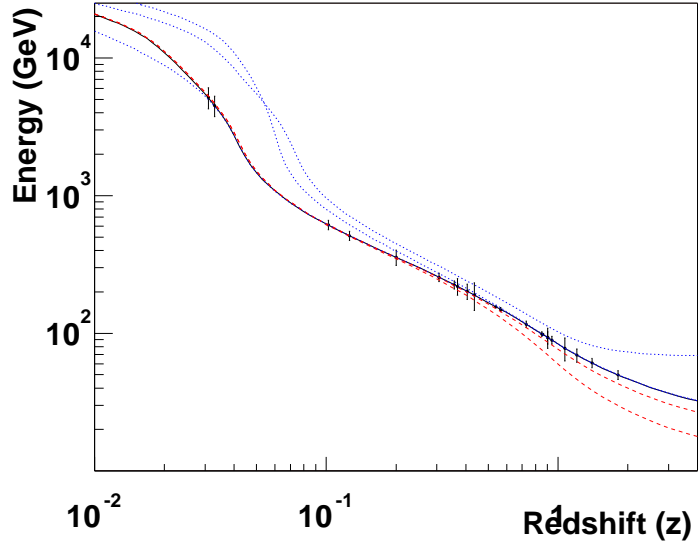}
  \end{center}
  \vspace{-0.5pc}
  \caption{GRH predictions as a function of redshift for several EBL
    models compared with the expected experimental accuracies of the
    GRH measurements for the different AGNs considered in this
    work. Blue dotted lines are for EBL models with low Star Formation
    Rate, warm dust and low IR background light. Red dashed lines only
    differ from the best fit model, which is the trend followed by the
    expected data points, by the amount of UV background light.}

\label{fig:GRH-models}
\end{figure}

\begin{itemize}
\item{} For most of the relevant parameters in the EBL prediction
(star formation rate, warm dust in the interstellar medium, IR
extragalactic background, ...) the most discriminating region is at
low redshift ($z > 0.1$) as can be seen in figure \ref{fig:GRH-models}.

\item{} The redshift evolution of the predicted GRH for the
different models in figure \ref{fig:GRH-models} shows that above redshift
0.1 the main parameter to take into account is the UV density (UV
model and high UV model).
\end{itemize}

Since most of the sensitivity to the $\Omega_M$ and
$\Omega_{\lambda}$ cosmological parameters is at large redshift
 (see fig.\ref{fig:models}), we've taken the following approach to
make a conservative estimate of the systematic uncertainties induced
by the present knowledge of the EBL:

\begin{itemize}
\item{} We have excluded the first two points (low redshift AGNs)
from the cosmological parameter fits assuming that they will be
primarily used to discriminate among different EBL modeling approaches.

\item{} We have used the remaining points (high redshift AGNs) to
fit the cosmological parameters. In order to quantify the additional
uncertainty due to the remaining EBL model dependence, we've
introduced in the fit as an additional parameter the amount of UV
background (using the same approach than in \cite{Kneiske03}.
\end{itemize}

\begin{figure}[h]
  \begin{center}
    \includegraphics[height=18pc]{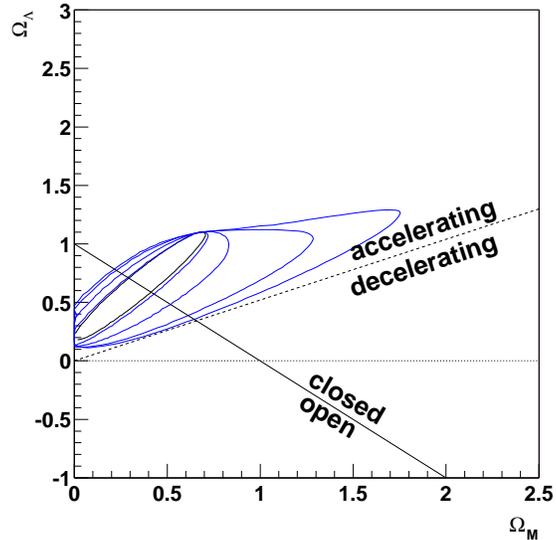}
  \end{center}
  \vspace{-0.5pc}
  \caption{Contour levels of $68 \%$ for the
    $\Omega_{M}-\Omega_{\lambda}$ plane for the fit with redshift $z>0.2$
    sources in which an additional constrained (at $5, 15, 25$ and $30
    \% $ levels)
    parameter has been introduced to account for the uncertainty in
    the UV background.}
    \label{UVconstrains}
\end{figure}

Unfortunately, this additional parameter turns out to be rather
correlated with the $\Omega_M$
and, fundamentally through this correlation, there is not hope to get
information on $\Omega_M$ and
$\Omega_{\lambda}$ without constraining the UV background. In figure
\ref{UVconstrains} it is shown how the contours in the $\Omega_M -
\Omega_{\lambda}$ degrade for different constrains. Assuming the
possibility to measure with an independent technique the UV background
at $15\%$ level, the quality of the Cosmological Parameter fits is still
very competitive.

At any rate, the launch of some new missions to study both the UV
\cite{UV-telescopes} background and the Star Evolution are
scheduled for the next few years, and they may shed new light into
our understanding of the EBL in the relevant energy region and
therefore help to substantially reduce the estimated systematical
uncertainties.

It is worth to point out here that, turning the other way around
the argument followed in this work, the mapping of the redshift
evolution of the GRH has been suggested already several times in
the literature as a way to constraint ONLY the EBL spectrum and
its redshift dependence provided that one assumes the constraints
in the cosmological parameters coming from other cosmology
measurements. In the strategy suggested in this work, it is
already suggested the use of the low-redshift GRH observations,
which are rather insensitive to the cosmological densities, to
constraint the low-redshift predictions of the EBL models. If one
really wants to use the gamma-ray absorption to measure values for
the EBL at different redshifts, the whole optical depth will be
needed. For that one needs to extract it from the spectrum
distortion knowing the original source spectrum, which may involve
source-dependent models for the extrapolation of the undistorted
spectrum. On the other hand, the approach discussed in this paper
relies on the hope that a more direct measurement of the EBL and
its redshift evolution in the relevant range becomes available in
the coming years.

\section{Conclusions}

Based on the extrapolation of the established extragalactic TeV
emitters and the best EGRET candidates to emit at the energy range
the new low-threshold IACTs such as the MAGIC Telescope will be
able to measure, the measurement  of the Gamma Ray Horizon for
sources in a large redshift range (0.031 to 1.8) has been
simulated. \par

It has been shown that percent level GRH measurements with
observation times of about 50 hours are expectable for some
sources, providing therefore a good mapping of the GRH as a
function of the redshift. These determinations will provide a new
distance estimator which has the following features:

\begin{itemize}

\item{}It is independent and behaves differently from the
luminosity-distance relation currently used by the Supernovae 1A
observations (see \cite{part1}).

\item{}It does not rely on the existence of a time-independent
standard-candle as do the Supernovae 1A measurements, although it
relies on the existence of a cosmological UV to infrared EBL which, in
first approximation, is assumed to be uniform and isotropic at
cosmological scales.

\item{}It uses Active Galactic Nuclei as sources, and therefore
may allow the study of the expansion of our universe up to the
highest observable redshifts. In this sense this method might
complement the picture provided by Supernovae 1A exploring the
farthest universe.

\end{itemize}

A multi-parameter fit of the cosmological parameters to $GRH(z)$ is
shown to provide at the statistical level a determination of
$\Omega_M$ v.s. $\Omega_\Lambda$ which is at the level of the
best present results from the combined Supernovae 1A observations.

In addition, the reduction of the energy threshold is expected to
allow the new instruments to discover a plethora of new sources
which has been hidden from our observation so far. An important
part of this new population is expected to be at relative large
redshift ($z>2$) where a big fraction of galaxies were AGNs and
hence could produce very high energy gamma-rays. These new sources
could drastically improve the results discussed here which,
therefore, have to be considered as being rather conservative.

In addition, a first estimation of the main systematic
uncertainties from experimental and theoretical origin has been
presented. The main experimental systematic has been estimated to
be the global energy scale of the IACTs which, as shown has a very
modest effect on the cosmological parameter fits.

The main theoretical systematic comes from the assumption of the UV to
infrared Extragalactic Background Light (EBL) and its redshift
evolution, for which several models have been explored in this
work. The systematic uncertainty coming from considering all these
models has been shown to be the dominant uncertainty in the
present situation.

These models have a quite broad spectrum of predictions in our
region of interest. There already exist which  some experimental data
able to constraint these models precisely in that region.
And, there is a good hope that the situation may
be better in the near future due to the lunch of
new missions to explore in detail the UV to infrared universe and to the
impressive improvement that the understanding of structure
formation in the universe is experimenting within the last few
years.

\section*{Acknowledgments}

We want to thank T.Kneiske and K.Mannheim for many discussions,
for providing us with their EBL spectra and for helping us to
cross-check some preliminary results of this work. We want to
thank our colleagues of the MAGIC collaboration for their comments
and support. We want to thank also G.Goldhaber, P.Nugent,
S.Perlmutter for their encouraging comments on the first phases of
this work. We are indebted to J. Garcia-Bellido for his very
valuable comments on this manuscript.





\end{document}